\documentclass[twocolumn,showpacs,preprintnumbers,amsmath,amssymb]{revtex4}

\usepackage{graphicx}% Include figure files
\usepackage{dcolumn}% Align table columns on decimal point
\usepackage{bm}% bold math
\usepackage{verbatim}% so you can comment large blocks of code
\usepackage{color}

\begin{document}

\title{The Maximum Patch Method for Directional Dark Matter Detection}

\author{Shawn Henderson}
 \email{shawnh@mit.edu}
\author{Jocelyn Monroe}
\affiliation{%
Department of Physics, Massachusetts Institute of Technology
}%
\author{Peter Fisher}%
\affiliation{%
Laboratory for Nuclear Science, MIT Kavli Institute for Astrophysics and Space Research, Department of Physics, Massachusetts Institute of Technology
}%

\date{July 22, 2008}

\begin{abstract}
Present and planned dark matter detection experiments search for 
WIMP-induced nuclear
recoils in poorly known background conditions.  In this environment, the
maximum gap statistical method provides a way of setting more sensitive
cross section upper limits by incorporating known signal information.  
We give a recipe for the numerical calculation of upper limits for 
planned directional dark matter detection experiments, that will measure 
both recoil energy and angle, based on the gaps between events in two-dimensional 
phase space.
\end{abstract}

\maketitle

%%%%%%%%%%%%%%%%%%%%%%%%%%%%%%%%%%%%%%%%%%%%%%%%%%%%%%%%%%%%%%%%%%%%%%%%%

\section{\label{sec:introduction}Introduction}
Dark matter comprises approximately 25\% of the mass of the
universe~\cite{r:spergel2003}.  The generic dark matter candidate is a
weakly interacting massive particle (WIMP).  If WIMPs are
supersymmetric particles, the predicted mass is in the range of 10 to
10$^4$ GeV/c$^2$, and the expected cross section lies in the range of
$10^{-42}$ to $10^{-48}$ cm$^2$~\cite{r:ellis2005}. Many experiments
seek to detect dark matter particles via their elastic scattering
interactions with detector nuclei~\cite{r:gaitskell2004}.  Recent
measurements~\cite{r:xenon2007,r:cdms2005} limit the cross section to
be less than approximately $5\times10^{-44}$ cm$^2$.
Given the small size of the expected WIMP cross section,
discrimination against backgrounds is of paramount importance in
direct detection experiments.  For the same reason, there is much to
gain by optimizing the statistical methods used to interpret
experimental data as upper limits on the WIMP interaction 
cross section~\cite{r:yellin}.

In this paper, we develop a new statistic, the maximum patch, for
setting limits on the WIMP-nucleus interaction 
cross section.  This method is motivated by directional dark matter
experiments, which seek to measure both the nuclear recoil energy, and the
recoil angle of the struck nucleon in WIMP-nucleon interactions.  In
section \ref{sec:overview} we introduce the theoretical distributions
used for setting limits, and in sections \ref{sec:1dstatistics} and
\ref{sec:2dstatistics} we discuss limit setting techniques within the
context of one- and two-dimensional WIMP detection experiments.  In
section \ref{sec:comparison} we compare results in the cases of (i)
observing a signal with no background, (ii) observing some signal and
some background, and (iii) observing only background.  

%%%%%%%%%%%%%%%%%%%%%%%%%%%%%%%%%%%%%%%%%%%%%%%%%%%%%%%%%%%%%%%%%%%%%%%%%

\section{\label{sec:overview}Setting Dark Matter Cross Section Limits}
Direct detection experiments typically measure the energy deposited by
the recoil nucleus~\cite{r:gaitskell2004}, infer the true nuclear
recoil energy, and set upper limits on the WIMP-nucleus interaction
cross section by comparing the theoretical distribution with this one-dimensional data set.  
The theoretical event rate distribution is given by~\cite{r:lewin}
\begin{equation}
\frac{dN}{dE} \ = \ \frac{R_0}{E_0 r} \frac{1}{2 \pi v_0^2} \int_{v_{threshold}}^{v_{max}} \frac{1}{v} f(v,v_E) d^3v
\label{eq:1Ddistribution}
\end{equation}
where $E$ is the nuclear recoil energy, $E_0=\frac{1}{2} m_D v_0^2$ 
is the dark matter particle's kinetic energy, $r = 4 m_D m_T / (m_D + m_T)^2$ 
with dark matter particle mass $m_D$ and target
nucleus mass $m_T$, $v_{threshold}$ and $v_{max}$ are the minimum observable 
and escape velocities of the dark matter (taken to be the dark
matter velocity that produces a maximum recoil energy equal to the experimental 
lower limit recoil energy threshold
and $\infty$ here respectively, for simplicity), $v_E$ = 244 km/s is the Earth's velocity
relative to the dark matter halo, and
$f(v,v_E)$ is the dark matter velocity distribution function, assumed
here to be a Maxwell-Boltzmann distribution with RMS velocity $v_{0}$ = 230 km/s.
The normalization factor $R_0$ is the event rate per unit mass
with $(v_{threshold},v_{max})$ = (0,$\infty$),
\begin{equation}
R_0 \ = \ \frac{2}{\sqrt{\pi}} \frac{N_0}{A} \frac{\rho_D}{m_D} v_0 \sigma_0
\label{eq:r0}
\end{equation}
where $N_0$ is Avogadro's number, $A$ is the atomic mass of the
target, $\rho_D$ is the dark matter density, taken here to be 0.3
$(GeV/c^2)/cm^3$, and $\sigma_0$ is the zero momentum-transfer dark
matter-nucleus interaction cross section.  We use the fact that 
the differential interaction rate scales simply with $\sigma_0$ in
the following discussion of limit setting techniques.

A new thrust in the field of WIMP searches has been to develop
detector sensitivity in a second dimension, the nuclear recoil
angle~\cite{r:Burgos:2007gv,r:Miuchi:2007ga,r:Dujmic:2007bd}.  The
WIMP-nucleus interaction signal is expected to be highly anisotropic
in recoil direction because of the earth's motion with respect to the
WIMP halo~\cite{r:Spergel:1987kx}.  In contrast, the backgrounds of
most WIMP experiments are relatively isotropic in recoil angle in the detector
coordinate system~\cite{r:mei2005}, and therefore this experimental approach can provide
increased discrimination against backgrounds.  It has recently been
suggested that WIMP direct detection searches relying on a recoil
energy signature alone may be insufficient for distinguishing WIMP
events from nuclear recoils if the WIMP-nucleus cross section is
smaller than the coherent scattering cross section for solar
neutrinos~\cite{r:jmonroe}.  This makes directional detection
particularly attractive, since solar neutrino-induced recoils point
back to the sun, unlike recoils from WIMP interactions.  Directional
detection could also potentially probe the velocity distribution of
our galaxy's dark matter halo~\cite{r:Host:2007fq}.  The theoretical
distribution as a function of nuclear recoil energy $E$ and recoil
angle $\psi$ (where $\psi$ is the angle in the detector lab frame between the nuclear 
recoil track observed in the detector and the direction the dark matter ``wind'' is blowing, 
which is normally taken to be the vector pointing from the constellation Cygnus to Earth) 
is given by~\cite{r:Spergel:1987kx,r:lewin}
\begin{equation}
\frac{d^2N(v_E,\infty)}{dE d(cos\psi)} \ = \ \frac{1}{2}\frac{R_0}{E_0 r}exp\biggl(-\frac{(v_E cos\psi - v_{min})^2}{v_0^2}\biggr)
\label{eq:2Ddistribution}
\end{equation}
where $v_{min} \ = \ (E / E_0 r)^{1/2} v_0$ is the smallest dark matter particle velocity which can produce a nuclear recoil with energy $E$.

Given a theoretical distribution, one can compare with an observation to
set a limit on the WIMP-nucleus interaction cross section.  The usual
method to obtain an upper limit at some confidence level is to vary
the theoretical parameters until the appropriate cumulative
probability distribution function (CDF) takes on the confidence level
desired ($0.9$ for a $90$\% confidence level upper limit) when
evaluated at the observed statistic (e.g. the number of observed
events).  In the following two sections, we discuss upper limit calculations using the one- 
and two-dimensional theoretical distributions respectively, together with several
statistics of interest.

%%%%%%%%%%%%%%%%%%%%%%%%%%%%%%%%%%%%%%%%%%%%%%%%%%%%%%%%%%%%%%%%%%%%%%%%%

\section{\label{sec:1dstatistics}Dark Matter Statistics in One Dimension}
Here we compare the traditional Poisson method with the maximum gap, a
statistic often used in dark matter experiments for obtaining an upper
limit with one-dimensional data~\cite{r:xenon2007,r:cdms2005}.  While
the traditional Poisson method is based solely on the number of counts
observed~\cite{r:pdg}, the maximum gap procedure incorporates what is
known about the shape of the expected signal into the limit
determination~\cite{r:yellin}.  For the discussion that follows,
consider a series of nuclear recoil energy measurements
$\{E_{1},...,E_{N}\}$ where $N$ is the total number of measurements.
Assume the data points are distributed with a
known theoretical function $dN(\vec{\lambda})/dE$, where the
$\vec{\lambda}$ are the parameters of the theoretical model $dN/dE$
given in equation~\ref{eq:1Ddistribution}.  Assuming standard values 
for the dark matter halo parameters, there is 
only one free parameter which we then vary to set upper limits, the 
zero-momentum transfer dark matter-nucleus interaction cross section
$\sigma_{0}$ in equation~\ref{eq:r0}.

%%%%%%%%%%%%%%%%%%%%%%%%%%%%%%%%%%%%%%%%%%%%%%%%%%%%%%%%%%%%%%%%%%%%%%%%%

\subsection{\label{subsec:poisson}Poisson Method}
A straightforward cross section upper limit on a given data set 
can be obtained by employing the Poisson method.  
To set a limit, we are interested in the probability, 
given a value of the cross section $\sigma_{0}$ in a theoretical
distribution $dN/dE$,
that the total number of events observed in our data
is equal to a certain value or less.  If we are conservative and assume no knowledge
of the background distribution and therefore that
all observed events are signal, then an upper limit at some desired confidence 
level may be set by adjusting $\sigma_{0}$ in $dN/dE$ until the total number of events $\mu$
expected, given by integrating $dN/dE$ over the 
whole experimental range, is such that it satisfies equation~\ref{eq:PoissonCDF}.
\begin{equation}
\alpha=e^{-\mu}\displaystyle\sum_{m=0}^N \frac{\mu^{m}}{m!}
\label{eq:PoissonCDF}
\end{equation}
Here, $1-\alpha$ is the confidence level of the upper limit set in this way, and $N$ is the
number of observed data events.  In order to incorporate knowledge of expected backgrounds 
into equation~\ref{eq:PoissonCDF}, we must use a modified form of this relation that assumes 
the overall normalization of the background, which is often poorly understood in dark matter 
direct detection experiments (for instance, see equation~32.35 in~\cite{r:pdg}).  
The most conservative approach is to assume no knowledge of the backgrounds, and so all
events are signal candidates.  In this case, any observed events
considerably degrade the upper limit obtained 
with equation~\ref{eq:PoissonCDF}.  This is particularly true for scenarios in which a background
fills a small subset of the full experimental acceptance.  For nuclear recoil signals, 
as the energy detection threshold is lowered, the sensitivity to 
backgrounds increases.  Background events observed near detection threshold are 
counted with the same significance as events in higher recoil energy sub-intervals of the 
experimental acceptance.  Direct dark matter detection experiments
gain sensitivity to WIMP events by lowering their energy thresholds, since the 
WIMP-nucleon interaction rate is expected to peak at low nuclear recoil energies.  In this 
scenario, the Poisson method can lead to overly conservative upper limits on the 
WIMP-nucleon interaction cross section, in the presence of backgrounds.

%%%%%%%%%%%%%%%%%%%%%%%%%%%%%%%%%%%%%%%%%%%%%%%%%%%%%%%%%%%%%%%%%%%%%%%%%

\subsection{\label{subsec:maxgap}Maximum Gap Method}
For a given $\sigma_{0}$, the ``gap''
for a pair of data points is defined to be~\cite{r:yellin}
\begin{equation}
x_{i}=\int^{E_{i+1}}_{E_{i}} \frac{dN}{dE}(\sigma_{0}) dE
\label{eq:gapDefinition}
\end{equation}
where $x_{i}$ is the value obtained by integrating $dN/dE$ between the
observed energy values $[E_{i},E_{i+1}]$ for $i=0,..,N$ (see
figure~1 in \cite{r:yellin}) and $E_{0}$ and $E_{N+1}$ are the lower
and upper recoil energy experimental thresholds.  
A set of $N$ recoil energy
measurements yields an $(N-1)$-dimensional vector $\vec{x}$ of gaps.
The ``maximum gap'' for a set of $N$ recoil energy measurements is
defined to be the largest member of the set of all gaps that can be
computed from the data.  This quantity depends on an integral over the
hypothesized theoretical distribution, and through that integral, on the
WIMP interaction cross section $\sigma_{0}$.  The larger the maximum
gap given a $\sigma_0$, the larger the discrepancy between the number of points observed
in data and the number of points expected.  Therefore, the maximum gap 
allows a powerful statistical test between the measured data and the
normalization of the theoretical signal distribution.  

To set a limit, we are interested in the probability, given a
value of the cross section $\sigma_{0}$ in a theoretical
distribution $dN/dE$, that the maximum gap for a given set of data is
equal to a certain value or less.  This is described by the CDF of the
maximum gap, and can be analytically calculated \cite{r:yellin},
\begin{equation}
C_{0}(x,\mu)=\displaystyle\sum_{k=0}^m \frac{(kx-\mu)^{k}e^{-kx}}{k!}\left(1+\frac{k}{\mu-kx}\right)
\label{eq:maximumGapCDF}
\end{equation}
where $\mu$ is the total number of events expected in the experimental
range ($dN/dE$ integrated from lower to upper limit energy threshold),
and $m$ is the greatest integer $\leq$ $\mu/x$.  An upper limit at a
given confidence level is obtained from
equation~\ref{eq:maximumGapCDF} by adjusting the input 
cross section $\sigma_{0}$ until the function $C_{0}$ above, evaluated at
the maximum gap of the data, yields $0.90$.  The interpretation 
is that in an ensemble of experiments, on each of
which the upper limit setting technique is employed, $90$\% will obtain
an upper limit greater than $\sigma_0$.
%
%%%%%%%%%%%%%%%%%%%%%%%%%%%%%%%%%%%%%%%%%%%%%%%%%%%%%%%%%%%%%%%%%%%%%%%%%

\subsection{\label{subsec:limittechniques}Discussion of Limit Setting Techniques}
The maximum gap statistic possesses a number of nice qualities,
particularly in the presence of background, which motivate the
generalization of the method to two dimensions for directional dark matter 
detection experiments.  We summarize the main
results here; see~\cite{r:yellin} for a rigorous discussion.

First, the maximum gap is unchanged under a one-one transformation of
the variable in which the events are distributed.  This can be seen 
by making a transformation from recoil energy in the above
discussion to a variable $\rho$ such that $\rho$ is equal to the total
number of events expected between the point $E$ and the lower energy
threshold.  In $\rho$, equation~\ref{eq:1Ddistribution} is a
uniformly distributed, unit density function.  This calculation is
treated in more detail in Appendix~I.  
This means that the maximum gap does not depend on the form of the theoretical
distribution.
Second, the method can be
used for an arbitrarily large number of observed data points, and
requires no binning of the data points.  Most importantly,
this statistic provides a conservative upper limit on the true WIMP
cross section that can considerably out-perform the Poisson upper limit
setting technique.

In WIMP detection experiments there are often low energy backgrounds
from processes which are difficult to model accurately with
simulations, such as MeV-scale neutron interactions or $^{238}$U and
$^{232}$Th decay progeny in the detector materials.
It is unlikely that the measured maximum gap will
be found in an interval of an experiment's acceptance that contains
both signal WIMP events and a large number of background events.
The presence of sizable background would significantly shorten the gap sizes
expected from signal alone.
We thus expect the maximum gap to occur in the data in an interval
where the background does not dominate.  In this way, the maximum gap method
automatically selects recoil energy intervals that are
characteristic of the expected WIMP signal alone.  

Another striking advantage of the maximum gap method is that, for a fixed gap
size, it is independent
of the total number of events observed.
In the Poisson case, each additional point
observed inflates the upper limit an experimenter sets on
their data.  On the other hand, if an experimenter observes a large gap in their data, the limit 
set on that data is unchanged if the number of points 
observed outside of the maximum gap is one or one million.  

%%%%%%%%%%%%%%%%%%%%%%%%%%%%%%%%%%%%%%%%%%%%%%%%%%%%%%%%%%%%%%%%%%%%%%%%%

\section{\label{sec:2dstatistics}A New Dark Matter Statistic for Two Dimensions}
For directional dark matter detection experiments, it is desirable to
preserve the benefits of the maximum gap method for setting upper
limits.  Towards this end, we need to generalize the method to two
dimensions.  We consider a series of measurements
$\{(E_{1},\cos(\psi)_{1}),...,(E_{N},\cos(\psi)_{N})\}$, where $N$ is
the total number of measurements of the energy of nuclear recoils $E$,
and $\psi$ is the nuclear recoil angle in a dark matter
detection experiment measured from the vector pointing from the constellation Cygnus
to the Earth in the detector lab frame.  In general, the two-dimensional rate will be
given by a function $d^{2}N(\vec{\lambda})/d(\cos(\psi))dE$,
where the $\vec{\lambda}$ are the theoretical parameters of the model.
As in one dimension, the model for the two-dimensional
differential WIMP-nucleon interaction rate, equation
\ref{eq:2Ddistribution}, under standard dark matter halo
assumptions, depends only on $\sigma_{0}$, the true WIMP interaction
cross section.  In the following we describe an algorithm for
obtaining general, multi-dimensional CDFs, focusing on the
two-dimensional case.  We then apply the usual prescription for setting
upper limits,
varying $\sigma_0$, and
find that our two-dimensional limit setting technique has the correct $90$\%
coverage.

%%%%%%%%%%%%%%%%%%%%%%%%%%%%%%%%%%%%%%%%%%%%%%%%%%%%%%%%%%%%%%%%%%%%%%%%%

\subsection{\label{sec:mc_cdf}Monte Carlo Generated Cumulative Distribution Functions}

To calculate the CDFs for an arbitrary statistic of interest (SI), we 
simplify matters by asking an easier
question than ``what is the probability that the SI 
is less than or equal to a certain value'' and instead ask ``what
is the probability that, given an observation of $N$ events, the SI is
less than or equal to a certain value.''  The advantage of the latter
question is that it can be addressed with Monte Carlo methods,
and leads naturally to the resolution of the first question.

We start by generalizing the one-dimensional case. 
The theoretical distribution in equation \ref{eq:1Ddistribution},
assuming a value for $\sigma_0$,
gives a concrete form for $dN/dE$, the expectation for how the
observed events are distributed in nuclear recoil energy.  Then,
we draw $N$ events from this distribution.  We compute the value of
the SI on this fake data, whether it be the maximum gap of the
distribution, or some other SI.
We repeat this procedure many times, until we have an
SI frequency distribution, given $N$ observed events.  We do this in
turn for $N=1,2,...,N_{max}$ stopping for $N_{max}$ so large that the Poisson
probability (equation~\ref{eq:PoissonProbability}) for observing $N_{max}$ 
events is negligible.  This results in
an array $\vec{h}=\{h_{0},...,h_{N_{max}}\}$ of histograms, 
where $h_{0}$ is the
frequency plot of the SI given the observation of zero events, $h_{1}$ 
is the frequency plot of the SI given the
observation of one event, etc.

In each of the $h_{i}$, $i=1,...,N_{max}$, we have $N_{toys}$ entries, where
$N_{toys}$ is the number of toy experiments, for each number
of observed events, that we performed.
By normalizing each of the $h_{i}$ by $N_{toys}$, we obtain the
probability distribution of the SI, for each of the $i$ events observed 
subclasses considered.  The resulting normalized vector of histograms 
can be properly interpreted as the probability distribution functions 
(PDFs) of the SI $\hat{h}=\vec{h}/N_{toys}$.

For setting an upper limit, we need the CDFs of the SI.  
To construct these, we take
each histogram member $\hat{h}_{i}$ of $\hat{h}$ in turn and create a
new histogram, $\hat{h}_{c,i}$,
assigning to each bin $b'_{mi}$ of $\hat{h}_{c,i}$ the value given 
in equation~\ref{eq:binToBinSum}, where $b_{ki}$ is the $k^{th}$ bin 
of $\hat{h}_{i}$, and $k$ and $m$ index the bins of the PDF and CDF histograms, 
respectively.
\begin{equation}
b'_{mi}=\displaystyle\sum_{k=1}^m b_{ki}
\label{eq:binToBinSum}
\end{equation}
For convenience, $500$ bin CDF and PDF histograms were generated for the
maximum gap studies in this paper, and $300$ bin CDF and PDF histograms
were generated for the maximum patch studies.

In this way we
numerically turn each of the PDFs in $\hat{h}$ into a CDF in
$\hat{h}_{c}$ for the SI.
Now we have $\hat{h}_{c}$, a vector of CDFs, each corresponding to a
number of observed events.  The probability of observing a
certain number of events is determined by the Poisson probability of observing
$N$ events given $\mu$ total events,
\begin{equation}
P(\mu,N)=\frac{\mu^{N}}{N!}e^{-\mu}.
\label{eq:PoissonProbability}
\end{equation}
In order to construct the full CDF of the SI for the input
theoretical distribution $dN(\sigma_{0})/dE$, we add
all of the histograms in $\hat{h}_{c}$ together, weighting each by the
probability in equation~\ref{eq:PoissonProbability} for seeing that
number of events.
This yields the following equation for the Monte Carlo generated CDF for
the SI,
\begin{equation}
C_{SI}(x_{SI},\mu)=\displaystyle\sum_{k=0}^{N_{max}} \hat{h}_{c,k}(x_{SI}) \left ( \frac{\mu^{k}}{k!}e^{-\mu} \right )
\label{eq:MCMaximumGapCDF}
\end{equation}
Here $x_{SI}$ is the value of the SI at which we want to know the
value of the CDF, and $\mu$ is the total number of events expected in the
experimental range.  The notation $\hat{h}_{c,k}(x_{SI})$ means to
evaluate the value of the $k^{th}$ histogram in the $\hat{h}_{c}$ vector of
SI CDFs at $x_{SI}$.  This can be done in a number of different ways;
we can take this as the height of the bin in $\hat{h}_{c,k}$ that
$x_{SI}$ falls into (thereby obtaining a discrete CDF), or, perform a 
bin-to-bin interpolation, thus
turning the $\hat{h}_{c,k}$'s into smooth functions of $x_{SI}$.  In
the results presented in this paper, the $\hat{h}_{c,k}$ histograms
are interpolated using splines.

We have now in equation~\ref{eq:MCMaximumGapCDF} 
manufactured the analogue to equation~\ref{eq:maximumGapCDF}
for the maximum gap statistic for an arbitrary SI.  Note however that
for the maximum gap statistic this discussion is unnecessarily
complicated because the maximum gap is unchanged under a one-one
transformation of the theoretical distribution $dN/dE$ in $E$, as
shown in Appendix~I.  This property allows one to
transform any distribution into a unit density, uniformly distributed
function.  Thus whenever we change $\sigma_{0}$ for the maximum gap
statistic, we need not draw events from a different distribution, we
may instead always use a unit density, uniform distribution.

We validate the Monte Carlo CDF generating scheme by comparing the
frequency distribution of upper limits resulting from the Monte
Carlo version of the maximum gap method with the result
obtained using the analytic CDF in equation~\ref{eq:maximumGapCDF}.  We
find that the results agree within numerical errors.  

%%%%%%%%%%%%%%%%%%%%%%%%%%%%%%%%%%%%%%%%%%%%%%%%%%%%%%%%%%%%%%%%%%%%%%%%%

\subsection{\label{sec:maxpatchmethod}Maximum Patch Method}
Analogously to the one-dimensional gap, we may construct a ``patch'' as a
subset of an experiment's total acceptance, which is determined by
$(E_{0},\cos(\psi)_{0})$, the energy and angle lower limit experimental threshold
of measurement, and $(E_{N+1},\cos(\psi)_{N+1})$, the energy and
angle upper limit experimental threshold of measurement.  We define a patch
for a set of $N$ data points to be
\begin{equation}
y_{ijk}=\int^{\cos(\psi)_{j}}_{\cos(\psi)_{k}} \int^{E_{i+1}}_{E_{i}} \frac{d^{2}N}{d(\cos(\psi))dE}(\sigma_{0}) dE d(\cos(\psi))
\label{eq:patchDefinition}
\end{equation}
where $i$ ranges from $0$ to $N$ and $j$ and $k$ range, independently,
from $1$ to $N$.  We require that $\cos(\psi)_{j}>\cos(\psi)_{k}$ and
that $E_{i}<E_{j}<E_{i+1}$ and $E_{i}<E_{k}<E_{i+1}$. 
We also include the additional patch candidate
not picked up by this prescription for every $i$; namely that one 
which has as its
borders in $\cos(\psi)$ the upper and lower angular limits
$[\cos(\psi)_{0},\cos(\psi)_{N+1}]$.
Equation~\ref{eq:patchDefinition} describes rectangular patches, whose
limits are $[E_{i},E_{i+1}]$ in $E$ and
$[\cos(\psi)_{k},\cos(\psi)_{j}]$ (plus
$[\cos(\psi)_{0},\cos(\psi)_{N+1}]$) in $\cos(\psi)$.  Some of the
rectangles described by equation~\ref{eq:patchDefinition} may have
points inside their boundaries; these are disqualified from being
maximum patches, for the same reason that gaps with points in them are
disqualified in the maximum gap method. In principle any two-dimensional 
shape can be used to
define a patch; we have chosen to use rectangles for ease of computation.  
It is possible that sensitivity could be gained
in this method by considering patch geometries other than rectangles.

\begin{figure}[htpb]
\vspace{-0.25cm}
\hspace{-1cm}
\centerline{\includegraphics[width=9cm]{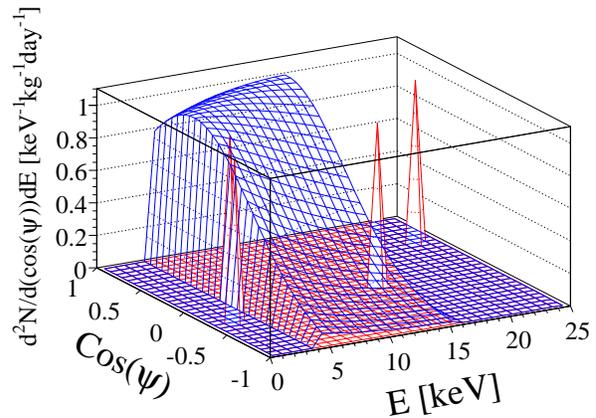}}
\vspace{-0.5cm}
\caption{\label{fig:method2D}An illustration of a maximum patch candidate.  
The spikes are hypothetical measured data points in an experiment, and 
the smooth curve is, for some assumed cross section, the expected distribution 
of signal between the two lowest energy data points.  The volume under the 
curve is one of the $y_{ijk}$'s of equation~\ref{eq:patchDefinition}
for this dataset, and thus a maximum patch candidate.  Note that if we project
the data points and theoretical distribution onto the energy interval, this 
is also a maximum gap candidate in the nuclear recoil energy dimension.}
\end{figure}

To set an upper limit we are interested in the maximum value that
$y_{ijk}$ takes in equation~\ref{eq:patchDefinition} for all
acceptable values of $j$, $k$ and $i$.  The situation is illustrated in
figure~\ref{fig:method2D}, where the spikes are Monte Carlo
generated fake data points in $E$ and $\cos(\psi)$ (4 in total) and
the smooth curve represents
$d^{2}N(\sigma_{0})/d(\cos(\psi))dE$ for a given cross section
value.  The maximum patch candidate in this
example has boundaries that extend from the lower to the upper
$\cos(\psi)$ threshold, and from $5-15$ $keV$.  Note that this
particular maximum patch candidate is also a maximum gap candidate in
$E$-space.

Our algorithm for computing the maximum patch on a set of observed
two-dimensional data points is described in detail in Appendix II.
\begin{figure}[htpb]
%\vspace{-0.25cm}
\centerline{\includegraphics[width=9cm]{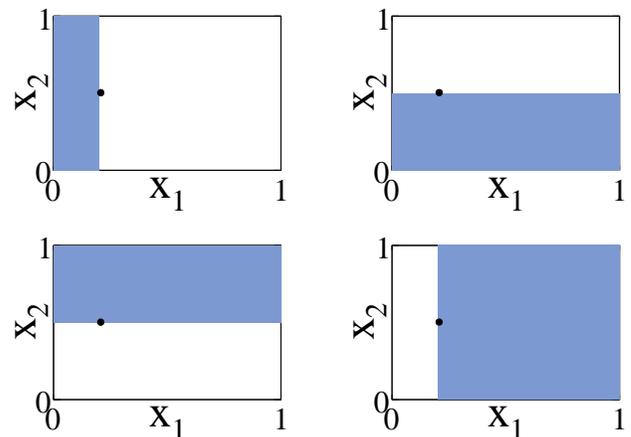}}
\vspace{-0.5cm}
\caption{\label{fig:maximumPatchAlgorithmExample}An example illustrating the 
maximum patch algorithm for 1 assumed data point in the experimental
acceptance.  If the data point is not on one of the four boundaries, there
are four patches that contribute.  Our algorithm is detailed in Appendix~II; 
the order in which our algorithm computes the above maximum patch candidates 
is upper left, upper right, lower left and lower right.}
\end{figure}
An example of the patch-finding algorithm for $1$ observed event is
shown in figure~\ref{fig:maximumPatchAlgorithmExample}.  Once the
maximum patch is found, one calculates the PDFs and sums them,
appropriately weighted, to produce the CDFs as in section~\ref{sec:mc_cdf}.
Figure~\ref{fig:maximumPatchCDFs} shows the resulting CDF for several
expected numbers of events $\mu$.  

We note that we are able to use an analogue 
to the simplified CDF generation scheme mentioned at the end of~\ref{sec:mc_cdf}
for the maximum gap method,
with one important change.  
Unlike the maximum gap statistic, the maximum patch is not unchanged 
under a one-one transformation of the theoretical distribution,
$d^{2}N/d(\cos(\psi))dE$, in $E$ and $\cos(\psi)$. 
Therefore, to set limits on data distributed according to $d^{2}N/d(\cos(\psi))dE$
with CDFs generated from a unit density, uniformly distributed function, one
must use the transformation given in Appendix C of~\cite{r:yellin2}.
\begin{figure}[htpb]
\vspace{-0.25cm}
\centerline{\includegraphics[width=9cm]{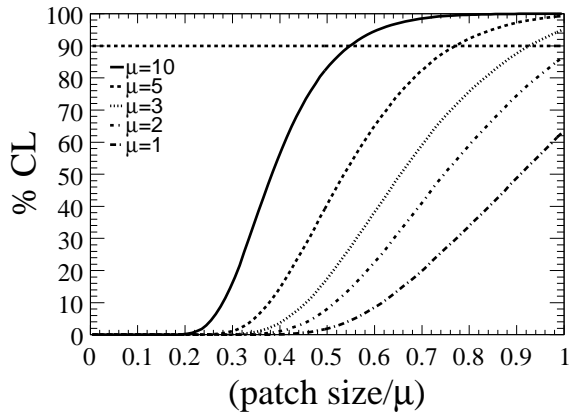}}
\vspace{-0.5cm}
\caption{\label{fig:maximumPatchCDFs}Cumulative probability distribution 
functions (CDFs) for various total expected numbers of events $\mu$ in the 
experimental acceptance (equation~\ref{eq:MCMaximumGapCDF} for the maximum 
patch SI, equation~\ref{eq:patchDefinition}).  A horizontal line is drawn at
the $90$\% confidence level.  So, for instance, for $\mu=5$, $90$\% of the 
time, the maximum patch of a toy signal experiment will be less than 
$\approx4$.}
\end{figure}

\subsection{\label{subsec:recipe}The Recipe}

The recipe for the experimenter wishing to use the maximum patch method to
set an upper limit on the dark matter cross section in her experiment is as
follows, 
assuming a measurement of a vector of $N$ two-dimensional data points
$\vec{D}=\{(E_{1},\cos(\psi)_{1}),...,(E_{N},\cos(\psi)_{N})\}$.  We take 
as an explicit example a direction sensitive dark matter direct detection 
experiment in this recipe, but this method can be used for any 
two-dimensional dataset for which the distribution of the signal is known.

\begin{enumerate}
\item Given a predicted two-dimensional rate $d^{2}N(\sigma_{0})/d(\cos(\psi))dE$, the experimenter calculates
the maximum patch of their data for some starting value $\sigma_1$ for the WIMP-nucleon 
interaction cross section
by applying the recipe for calculating the maximum patch of a set of two-dimensional 
data points outlined in Appendix~II.
\item The experimenter then must evaluate equation~\ref{eq:MCMaximumGapCDF} for the
case where the statistic of interest is the maximum patch.  The maximum patch CDF for $\mu$ 
total expected events can either be calculated by following the Monte Carlo procedure outlined 
in section~\ref{sec:mc_cdf} or by referencing the tables provided 
at the end of this paper in Appendix~III.  

\item Evaluating equation~\ref{eq:MCMaximumGapCDF} at the maximum patch
of the data yields the confidence level at which
$\sigma_1$ is an upper limit on the WIMP-nucleon interaction cross section.  If this is less (more) than
the confidence level desired, the cross section guess is increased (decreased) to a new guess $\sigma_2$.
The experimenter again calculates the maximum patch of the data for this assumed cross section and 
then, via equation~\ref{eq:MCMaximumGapCDF}, calculates the CL at which $\sigma_2$ is an upper limit 
on the WIMP-nucleon interaction cross section.  This procedure is iterated for as many guesses 
$\sigma_N$ as required to set the desired confidence level upper limit on the WIMP-nucleon 
cross section for the data.
\end{enumerate}

%%%%%%%%%%%%%%%%%%%%%%%%%%%%%%%%%%%%%%%%%%%%%%%%%%%%%%%%%%%%%%%%%%%%%%%%%

\subsection{\label{subsec:validation}Validation of the Maximum Patch Method}
\begin{figure}[htpb]
\vspace{-0.25cm}
\centerline{\includegraphics[width=9cm]{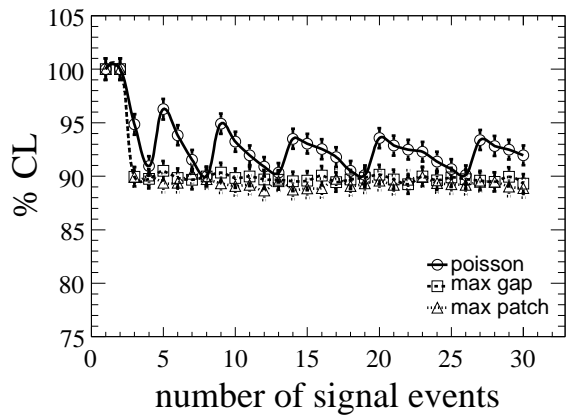}}
\vspace{-0.5cm}
\caption{\label{fig:upperLimitCoveragesPlot} For toy Monte Carlo experiments with pure 
signal events, the coverage as a function of signal events for the maximum gap, maximum patch
and Poisson methods.  This demonstrates that our 
upper limit setting methods have the correct coverage, within statistical errors for 
$\mu>2.23026$, as expected.  Each point in this plot corresponds to 10,000 MC toy experiments.
The errors shown are statistical.}
\end{figure}

We validate the maximum patch prescription described above by checking
that the coverage of our upper limit setting technique is correct.  To
do this, we generate many ensembles of toy Monte Carlo experiments, each with different
WIMP-nucleon interaction cross sections, and with signal events
distributed according to equation \ref{eq:2Ddistribution}.  We set an upper limit on each toy 
experiment using the maximum patch technique.  
Note that to test coverage at the $90$\% confidence level, we must choose cross sections
that yield an expected number of events $\mu>2.3026$.  Below this, no cross section 
upper limit can be set with a confidence level as high as $90$\% 
(see figure~\ref{fig:maximumPatchCDFs}).
For each input cross section $\sigma_0$, and the associated total number of 
events $\mu$, we perform 10,000 toy experiments, where the observed
number of events is Poisson-distributed 
about $\mu$, and
$(E,\cos(\psi))$ for the events are distributed according to equation~\ref{eq:2Ddistribution}.  
For each ensemble of 10,000 experiments with different input $\sigma_0$'s,
we record the percentage of the time our upper limit is higher than $\sigma_0$.  For 
an upper limit requested at $90$\% CL, the percentage should be $90$\% within 
statistical errors, which is the definition of correct coverage.  The results of this study, for 
$\sigma_0$'s such that $\mu=1,...,30$ 
are shown in figure~\ref{fig:upperLimitCoveragesPlot}.  
The stepping behaviour observed in the Poisson limit coverage is expected, due to the 
discrete nature of the statistic.
Figure~\ref{fig:upperLimitCoveragesPlot}
also shows the coverage computed in this way for the maximum gap method (with the CDFs
computed via our Monte Carlo technique) and the Poisson method.  All methods are observed to
have the correct coverage, within statistical and numerical errors.
%

%%%%%%%%%%%%%%%%%%%%%%%%%%%%%%%%%%%%%%%%%%%%%%%%%%%%%%%%%%%%%%%%%%%%%%%%%

\section{\label{sec:comparison}Comparison of Methods}
Having built the maximum patch method, we compare it to various other
methods for setting upper limits, in several circumstances of interest. 
Our goal is to highlight the impact of directionality for dark matter direct 
detection experiments.  Unless otherwise stated, for the various comparisons in this paper 
we arbitrarily assume a WIMP mass of $60$ GeV, and we use the Xenon10 
experiment's acceptance and target properties~\cite{r:xenon2007} to construct limits 
(ignoring subtleties like quenching factors, form factors and efficiencies). 

\begin{figure}[htpb]
\vspace{-0.25cm}
\centerline{\includegraphics[width=9cm]{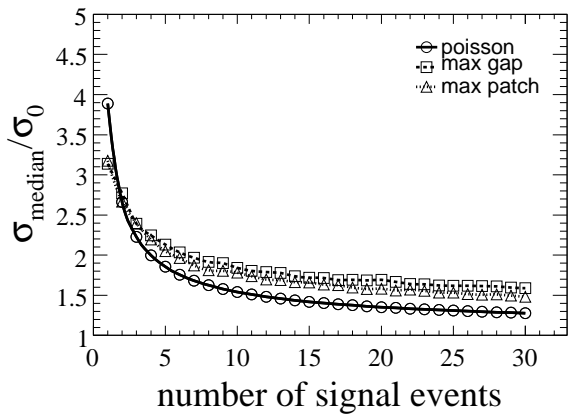}}
\vspace{-0.5cm}
\caption{\label{fig:upperLimitFrequencyMedians}The median upper limit cross section 
obtained by our implementation of the Poisson, maximum gap and maximum patch procedures 
divided by the true input cross section $\sigma_{0}$ as a function of input cross section 
(and thereby, as a function of the total number of expected events in the experimental
interval).  Each point in this plot corresponds to 10,000 MC toy experiments.}
\end{figure}

\begin{figure}[htpb]
\vspace{-0.25cm}
\centerline{\includegraphics[width=9cm]{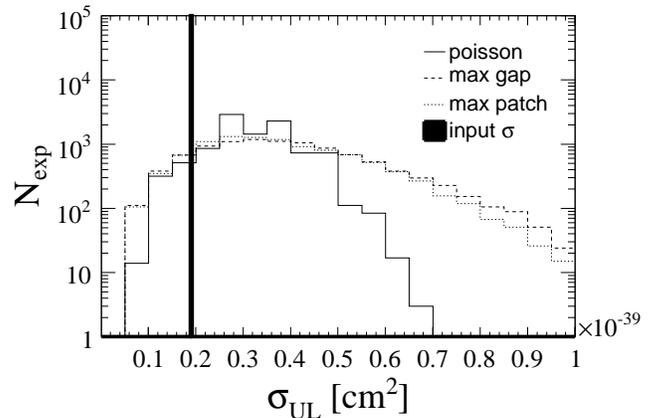}}
\vspace{-0.5cm}
\caption{\label{fig:upperLimitFrequencyDist}The frequency distribution of upper limits $\sigma_{UL}$ obtained by applying the
Poisson, maximum gap and maximum patch procedures to 10,000 toy Monte Carlo experiments with an input cross section $\sigma_{0}$ 
shown as a black line on the graph.  $\sigma_{0}$ was chosen to yield $7$ total expected events in the experimental 
interval.  The percent of the time that our procedure for each method sets a correct upper limit (a limit above the true 
input cross section $\sigma_0$) is $(90\pm1)$\%, $(90\pm1)$\% and $(92\pm1)$\% for the maximum gap, maximum patch and Poisson methods, 
respectively.}
\end{figure}

First, in the absence of background and with a sizable signal, we
would like to verify that the maximum patch method has not only the same
coverage as the  Poisson method (see subsection~\ref{subsec:validation}) but 
also that its performance is comparable as a method for setting upper limits.  
Towards this end, we employ the ensembles of 10,000 toy Monte Carlo experiments
from subsection~\ref{subsec:validation}, recording the median upper limit
obtained by the maximum patch, maximum gap and Poisson methods as a function
of $\mu$ for each 10,000 event sub-sample generated with a different input
$\sigma_0$.  This comparison is shown in figure~\ref{fig:upperLimitFrequencyMedians}.
Figure~\ref{fig:upperLimitFrequencyDist} shows the frequency distribution of upper limits
set by the maximum gap, maximum patch and Poisson methods on the 10,000 event ensemble with an input
$\sigma_0$ such that a total of $\mu=7$ signal events are expected.  These three distributions are
used to generate the $\mu=7$ point in both figure~\ref{fig:upperLimitFrequencyMedians} and 
figure~\ref{fig:upperLimitCoveragesPlot}.  The computed coverages in figure~\ref{fig:upperLimitFrequencyDist}
for the maximum gap, maximum patch and Poisson methods are $(90\pm1)$\%, $(90\pm1)$\% and $(92\pm1)$\%, 
respectively, which is correct, within statistical errors.

We note that in the case of pure signal, the Poisson method outperforms the 
maximum gap and maximum patch methods by a factor of $\approx1.2$.  This is 
expected; the maximum gap procedure has already been demonstrated to give 
looser upper limits than the Poisson method in the case of pure signal  
(see~\cite{r:yellin}, figure~3(a)).  This could be resolved, as in~\cite{r:yellin} 
by considering gaps containing greater than zero events, which is termed 
the optimal gap method.
The extension of the optimal gap approach to two dimensions is not considered here.
We also note that little
sensitivity is gained by using the maximum patch method versus the maximum gap 
method in the case of pure signal.  

\begin{figure}[htpb]
\vspace{-0.25cm}
\centerline{\includegraphics[width=9cm]{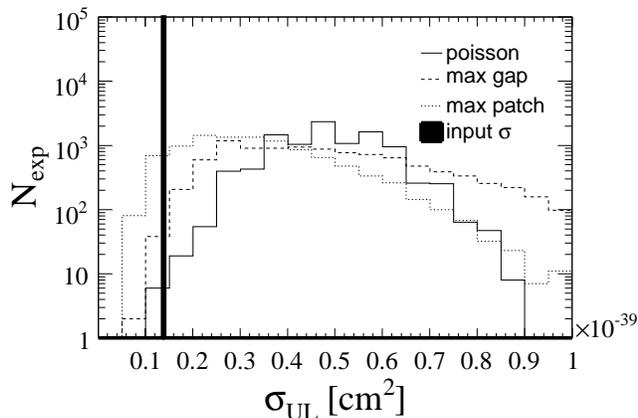}}
\vspace{-0.5cm}
\caption{\label{fig:upperLimitFrequencyDist2}The frequency distribution of upper limits $\sigma_{UL}$ obtained by applying the
Poisson, maximum gap and maximum patch procedures to 10,000 toy Monte Carlo experiments with an input cross section $\sigma_{0}$ 
shown as a black line on the graph.  $\sigma_{0}$ was chosen to yield $5$ total expected signal events in the experimental 
interval.  Background events were included in these experiments with a uniform distribution in the lower half of the recoil energy and 
recoil angle experimental acceptance.  For this plot, the total number of expected background events was set at $7$.  
The percent of the time that our procedure for each method sets a correct upper limit (a limit above the true 
input cross section $\sigma_0$) is $(100\pm1)$\%, $(95\pm1)$\% and $(100\pm1)$\% for the maximum gap, maximum patch and Poisson methods, 
respectively.}
\end{figure}

\begin{figure}[htpb]
\vspace{-0.25cm}
\centerline{\includegraphics[width=9cm]{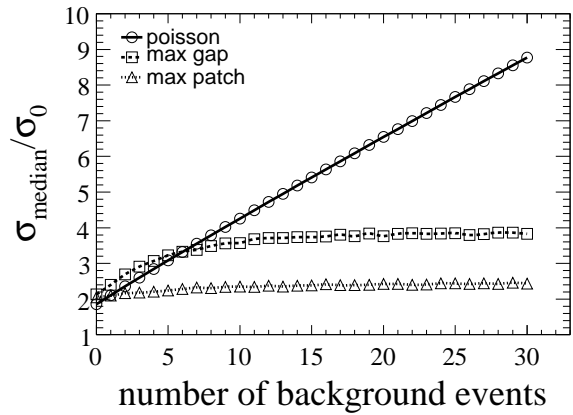}}
\vspace{-0.5cm}
\caption{\label{fig:upperLimitFrequencyMedians2}The median upper limit cross section obtained by our implementation of 
the Poisson, maximum gap and maximum patch procedures divided by the true input cross section $\sigma_{0}$ as a function
of background events injected into our toy experiments, uniformly distributed in the lower half recoil energy and recoil angle 
interval.  The point at number of background events equals $7$ comes from dividing the medians of the frequency distributions in 
figure~\ref{fig:upperLimitFrequencyDist2} by the true input cross section for the Poisson, maximum gap and maximum patch limit 
setting procedures.  Each point in this plot corresponds to 10,000 MC toy experiments and has, in addition to background, a signal 
component generated with a cross section corresponding to $5$ total expected signal events.}
\end{figure}

Realistically, WIMP direct detection experiments have
backgrounds, and so a more interesting comparison is how the maximum
gap, maximum patch and Poisson methods do in the presence of sizable
backgrounds.  For this test, we populate the lower half of our nuclear
recoil energy range, and the lower half of our nuclear recoil angular range,
with a background drawn according to a flat distribution.  
We caution that all of our results including background events are highly dependent on 
this particular background distribution choice.
Figure~\ref{fig:upperLimitFrequencyDist2} shows
the frequency distribution of upper limits obtained by applying the
maximum gap method, the maximum patch method, and the Poisson method
to 10,000 toy experiments generated in this way with a total number of 
expected background events of $7$, and a total expected number of WIMP signal 
events of $5$.  The coverage is
$(100\pm1)$\%, $(95\pm1)$\% and $(100\pm1)$\% for the maximum gap, maximum patch
and Poisson methods respectively, which
is not at the confidence level requested due to the
presence of the large background.  The median $90$\% confidence level upper
limit cross sections, from the maximum gap, maximum patch and Poisson
techniques are compared in
figure~\ref{fig:upperLimitFrequencyMedians2}, as a function of the
total number of expected input background events $\mu=1,2,...,30$ (with
$5$ expected signal events in each toy experiment).  The total number of background events
in a given Monte Carlo experiment, like the total number of signal events, is drawn randomly from a Poisson
distribution with the mean given by the total number of expected events.

Figure~\ref{fig:upperLimitFrequencyMedians2} shows that in the presence of a sizable WIMP signal, the maximum 
patch procedure provides stronger upper limits than the Poisson or maximum gap procedures as the 
amount of background contamination increases.  The Poisson method does so poorly because
it uses only the total number of events to set upper limits,
assuming that they are all signal, yielding an increasingly inflated upper limit as
the number of background events injected into the toy experiments increases.  The maximum
patch method outperforms the maximum gap procedure because it includes an
extra dimension in which signal and background are differently distributed.  The observation that the maximum gap and maximum patch limits seem to 
asymptotically flatten is due to the overly simplistic background chosen for these studies; 
eventually the maximum patch or gap is always outside of the lower $E$ interval or 
$E-\cos(\psi)$ quadrant, and thus characteristic only of the WIMP-signal input which, within
statistical fluctuations, is identical outside of the lower $E$ interval or $E-\cos(\psi)$ 
quadrant as the backgrounds are confined there.  

\begin{figure}[htpb]
\vspace{-0.25cm}
\centerline{\includegraphics[width=9cm]{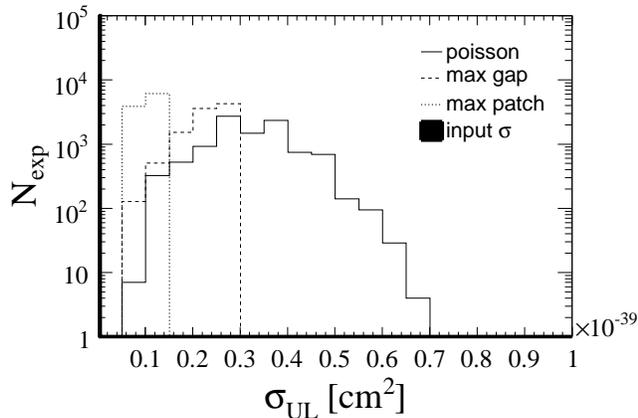}}
\vspace{-0.5cm}
\caption{\label{fig:upperLimitFrequencyDist3}The frequency distribution of upper limits $\sigma_{UL}$ obtained by applying the
Poisson, maximum gap and maximum patch procedures to 10,000 toy Monte Carlo experiments with an input cross section 
$\sigma_{0}=1\times10^{-46}$ 
shown as a black line on the graph, chosen to give $\approx0$ events for the Monte Carlo toy exposures.  
Background events were included in these experiments with a uniform distribution in the lower half of the recoil energy and 
recoil angle experimental acceptance.  For this plot, the total number of expected background events was set at $7$.  
The percent of the time that our procedure for each method sets a correct upper limit (a limit above the true 
input cross section $\sigma_0$) is $(100\pm1)$\%, $(100\pm1)$\% and $(100\pm1)$\% for the maximum gap, maximum patch and Poisson methods, 
respectively.}
\end{figure}

\begin{figure}[htpb]
\vspace{-0.25cm}
\centerline{\includegraphics[width=9cm]{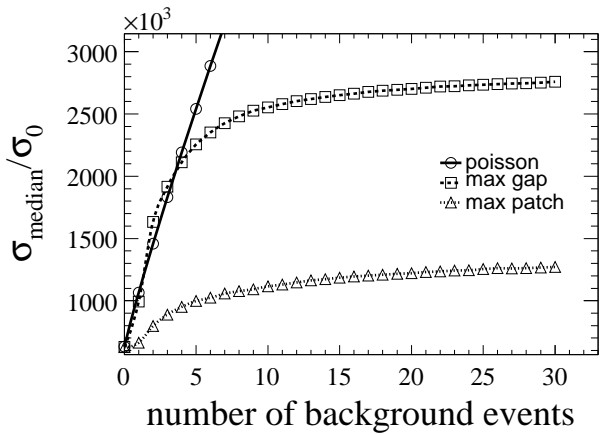}}
\vspace{-0.5cm}
\caption{\label{fig:upperLimitFrequencyMedians3}The median upper limit cross section obtained by our implementation of 
the Poisson, maximum gap and maximum patch procedures divided by the true input cross section $\sigma_{0}$ as a function
of background events injected into our toy experiments, uniformly distributed in the lower half recoil energy and recoil angle 
interval for a negligible input signal cross section.  The point at number of background events equals $7$ comes from dividing the 
medians of the frequency distributions in 
figure~\ref{fig:upperLimitFrequencyDist3} by the true input cross section for the Poisson, maximum gap and maximum patch limit 
setting procedures.  Each point in this plot corresponds to 10,000 MC toy experiments.}
\end{figure}

The most probable situation in current dark matter experiments is that
the true cross section lies well below the detectable range.
To study this scenario, we perform 10,000 toy
experiments as above, with $\sigma_0=1\times10^{-46}$ cm$^2$ (or $\approx0$ total expected signal events), and the same
distributions of background events as in figures~\ref{fig:upperLimitFrequencyDist2} and 
~\ref{fig:upperLimitFrequencyMedians2} (flat in the $E-\cos(\psi)$ plane, and confined to the lower $E-\cos(\psi)$
quadrant).
Figure~\ref{fig:upperLimitFrequencyDist3} shows the frequency
distribution of upper limits obtained by applying the maximum gap
method, the maximum patch method, and the Poisson method to 10,000 toy experiments with an 
expected number of background events of $7$ and negligible signal.  The
coverage is $(100\pm1)$\%, $(100\pm1)$\% and $(100\pm1)$\% for the maximum gap, maximum patch and Poisson methods,
respectively, which is not at the confidence level
requested due to the presence of a large background and no signal.  The median $90$\%
confidence level upper limit cross sections, from the maximum gap,
maximum patch and Poisson techniques are compared in
figure~\ref{fig:upperLimitFrequencyMedians3} as a function of input background events in the case
of negligible signal.

In the presence of a negligible WIMP signal and increasing backgrounds, the maximum 
patch procedure provides by far the most restrictive upper limit as the 
amount of background contamination increases.  From figure~\ref{fig:upperLimitFrequencyMedians3} it is clear
that the Poisson method limit is not competitive as the number of backgrounds increases, and the maximum patch
method outperforms the maximum gap method by at least a factor of $2$ for more than $1$ expected background
events for this particular background distribution.


\begin{figure}[htpb]
\vspace{-0.25cm}
\centerline{\includegraphics[width=9cm]{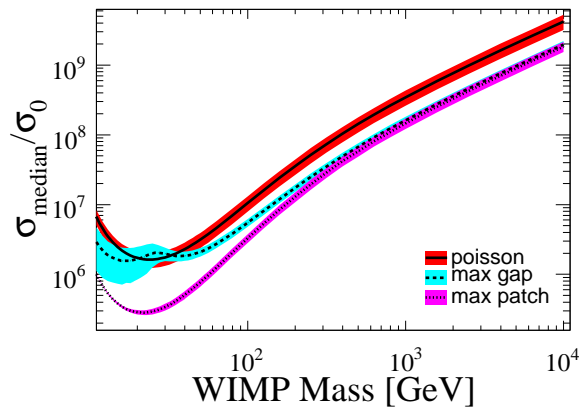}}
\vspace{-0.5cm}
\caption{\label{fig:upperLimitFrequencyMediansPlotVaryingMdark}The median upper limit cross section obtained by our implementation of 
the Poisson, maximum gap and maximum patch procedures divided by the true input cross section $\sigma_{0}$ as a function
of assumed WIMP mass, with $10$ background events on average that are uniformly distributed in the lower half recoil energy and recoil angle 
interval for a negligible input signal cross section.  Each point in this plot corresponds to 10,000 MC toy experiments.  The
bands correspond to the RMS widths of the upper limit frequency distributions obtained using the Poisson, maximum gap and maximum patch methods, 
respectively, at each mass point sampled.}
\end{figure}

In typical dark matter detection experiments, upper limits on the WIMP-nucleon cross section are reported 
as a function of WIMP mass.
In order to compare the performance of the maximum gap, maximum patch and Poisson methods of
setting upper limits as a function of WIMP mass, we generated 10,000 Monte Carlo toy datasets at
a variety of different WIMP masses.  Adopting the likely scenario for a given dark matter
experiment, we hypothesize a WIMP-nucleon interaction 
cross section of $\sigma_0=1\times10^{-46}$ for these toy experiments and a background with an average number of events equal
to $10$ (again, flat in the $E-\cos(\psi)$ plane, and confined to the lower $E-\cos(\psi)$
quadrant).  The result of this study is the three limit curves in figure~\ref{fig:upperLimitFrequencyMediansPlotVaryingMdark}.
The solid bands represent the RMS widths of the upper limit frequency plots (much like 
figures~\ref{fig:upperLimitFrequencyDist3},~\ref{fig:upperLimitFrequencyDist2} and~\ref{fig:upperLimitFrequencyDist})
obtained on the 10,000 experiments at each given mass point.
The shape of the maximum gap limits at low WIMP masses is an artifact of the background choice.
For low WIMP masses, the maximum patch method is by far the most sensitive to the cross section (between 
$\approx10-100$ GeV)
with a very small RMS.  At higher WIMP masses, the difference between the maximum gap and maximum patch limit
techniques diminishes, but both consistently outperform the Poisson limit setting method.  We note that
the studies in this paper are all based on one assumed background shape.  Different background distributions
will lead to different results.
%

%
%%%%%%%%%%%%%%%%%%%%%%%%%%%%%%%%%%%%%%%%%%%%%%%%%%%%%%%%%%%%%%%%%%%%%%%%%

\section{\label{sec:conclusions}Conclusions}

In this paper, we have developed a new method for setting upper limits in two dimensions.  
The motivation for our maximum patch method is directional dark matter detection, but it is 
generally applicable to any two-dimensional data sets for which the distribution of the 
signal is known.  The approach is an extension of the one-dimensional maximum gap 
method~\cite{r:yellin}, a statistic
often used to set limits on the WIMP-nucleon interaction cross section in
direct detection dark matter experiments.  To directly detect dark matter
requires unprecedented control and understanding of backgrounds.  The great
advantage of the maximum gap and patch methods is that they require no knowlege
of the background distribution to set conservative upper limits on the WIMP
nucleon scattering cross section.  The scattering kinematics of a dark matter
signal in one and two-dimensional direct detection experiments are relatively
simple.  This information is included in a straightforward way in the maximum
gap and patch methods.  In particular, the maximum patch method incorporates
information about the large expected angular anisotropy of dark matter
scattering into the limit setting procedure.  We demonstrate that for
simplistic background assumptions, the maximum patch method and two-dimensional
dark matter detection yield a large gain in sensitivity, especially at low
WIMP masses, over one-dimensional dark matter detection.

%
%%%%%%%%%%%%%%%%%%%%%%%%%%%%%%%%%%%%%%%%%%%%%%%%%%%%%%%%%%%%%%%%%%%%%%%%%

\section*{\label{sec:app1}Appendix I}
Our goal is to prove, in detail, the assertion in \cite{r:yellin} that
the maximum gap is unchanged under a one-one transformation of the
variable in which the events are distributed, and thus that the
maximum gap is independent of the particular way in which the events
are distributed for a given $\sigma_{0}$ in one dimension (by
$dN(\sigma_{0})/dE$).  The total number of events in the experimental
window is given by $\mu$, where
\begin{equation}
\mu=\int^{+\infty}_{-\infty} dE \frac{dN}{dE}.
\label{eq:appendix1}
\end{equation}
Here, if $E$ is outside of the experimental thresholds,
$dN/dE=0$, and $dN/dE$ is strictly positive.

We wish to change variables from $E$ in equation~\ref{eq:appendix1}
to $\rho$, where $\rho$ \cite{r:yellin} is
\begin{equation}
\rho(E)=\int^{E}_{-\infty} dE'\frac{dN}{dE'}
\label{eq:appendix2}
\end{equation}
By the chain rule,
\begin{equation}
\frac{dN}{d\rho}=\frac{dN}{dE}\frac{dE}{d\rho}
\label{eq:appendix3}
\end{equation}
and $dE/d\rho$=$(d\rho/dE)^{-1}$.  $d\rho/dE$ from equation~\ref{eq:appendix2} is
then given by the fundamental theorem of calculus to be
\begin{equation}
\frac{d\rho}{dE}=\frac{d}{dE}\left(\int^{E}_{-\infty} dE'\frac{dN}{dE'}\right)=\frac{dN}{dE}(E)
\label{eq:appendix4}
\end{equation}
Thus, equation~\ref{eq:appendix3} implies that 
\begin{equation}
\frac{dN}{d\rho}=\frac{dN}{dE}\left(\frac{d\rho}{dE}\right)^{-1}=\frac{\left(\frac{dN}{dE}\right)}{\left(\frac{dN}{dE}\right)}=1
\label{eq:appendix5}
\end{equation}
and that in the new variable $\rho$, the differential rate is a
unit density function, of length $\mu$.

%%%%%%%%%%%%%%%%%%%%%%%%%%%%%%%%%%%%%%%%%%%%%%%%%%%%%%%%%%%%%%%%%%%%%%%%%

\section*{\label{sec:app2}Appendix II}
Our method for calculating the maximum patch is as follows.  A
similar scheme is put forth in \cite{r:yellin2}.  Call the
set of measured two-dimensional data points on which we wish to 
determine the maximum patch, $\vec{D}=\{(E_{1},\cos(\psi)_{1}),...,(E_{N},\cos(\psi)_{N})\}$, where 
as above, $(E_{0},\cos(\psi)_{0})$ and $(E_{N+1},\cos(\psi)_{N+1})$ we define to be
the upper and lower recoil energy and recoil angle thresholds of our experiment.

\begin{enumerate}
	\item Order $\vec{D}$ in $E$.  Call the new, $E$-ordered
	vector $\vec{D}'$.  \item Loop through the intervals
	$[E_{i},E_{j}]$ in $\vec{D}'$ such that $i$ runs from $0$ to
	$N+1$ and $j$ runs from $i+1$ to $N+1$.  \item Inside each $E$
	interval in this loop, if there are no points with an $E_{k}$
	such that $E_{i}<E_{k}<E_{j}$, then compute
	equation~\ref{eq:patchDefinition} with $E$ limits
	$[E_{i},E_{j}]$ and $\cos(\psi)$ limits
	$[\cos(\psi)_{0},\cos(\psi)_{N+1}]$.  \item If there are
	$N_{p}$ points with an $E_{k}$ such that $E_{i}<E_{k}<E_{j}$,
	make a list of them,
	$\vec{P}=\{(E_{0},\cos(\psi)_{0}),...,(E_{k},\cos(\psi)_{k}),...,$
	$(E_{N+1},\cos(\psi)_{N+1})\}$
	of length $N_{p}+2$, ordered in $E$, adding the points
	$(E_{0},\cos(\psi)_{0})$ and $(E_{N+1},\cos(\psi)_{N+1})$ (the
	threshold points - that's where the $+2$ comes from in the
	total number of points in $\vec{P}$) to the front, and back of
	the list, respectively.  Order $\vec{P}$ by $E$, calling the
	new $E$-ordered vector $\vec{P'}$.  Then loop through the
	intervals $[\cos(\psi)_{P',m},\cos(\psi)_{P',n}]$, in
	$\vec{P'}$ such that $m$ runs from $0$ to $N_{p}+2$ and $n$
	runs from $m+1$ to $N_{p}+2$, where the subscript $(P',m)$
	denotes the $m^{th}$ member of the ordered vector $\vec{P'}$.  Each
	iteration of this loop will provide a maximum patch candidate
	with $E$ limits $[E_{i},E_{j}]$ and $\cos(\psi)$ limits
	$[\cos(\psi)_{P',m},\cos(\psi)_{P',n}]$.  \item For each of
	these candidates, check to see if there is a point $D'_{d}$ in
	$\vec{D'}$ such that $E_{i}<E_{d}<E_{j}$ and
	$\cos(\psi)_{P',m}<\cos(\psi)_{d}<\cos(\psi)_{P',n}$.  If so,
	then there is a point inside our patch candidate, which means
	it is not a maximum patch candidate; throw it out.  \item If a
	maximum patch candidate has passed all of the above criterion,
	stick it in a vector of some arbitrary length $\vec{y}$.  This
	vector exhausts all possible rectangles in the two-dimensional 
	$E$-$\cos(\psi)$ plane.  \item To find the maximum
	patch of the data, loop through all of the elements $y_{i}$ of
	$\vec{y}$, and store the largest $y_{i}$ value; this is the
	maximum patch.
\end{enumerate}

%%%%%%%%%%%%%%%%%%%%%%%%%%%%%%%%%%%%%%%%%%%%%%%%%%%%%%%%%%%%%%%%%%%%%%%%%

\section*{\label{sec:app3}Appendix III}

Tables~I and~II below record the $\hat{h}_{c,k}(x_{SI})$'s of the maximum patch statistic for $k=1,...,100$ ($\hat{h}_{c,0}(x_{SI})$
is $0$ for all maximum patch values except for the patch equal to the total number of expected events).
The maximum patch CDF is computed by interpolating these points into smooth curves and using them to evaluate 
equation~\ref{eq:MCMaximumGapCDF}.  Note that the smoothed curves obtained from the tables are functions of $(y/\mu)$, 
the maximum patch value divided by the total number of expected events.  The work in this paper was done using 
$200$ CDFs in the sum of equation~\ref{eq:MCMaximumGapCDF}.  In Tables~I and~II, a deviation from a cumulative probability
of $100$\% for $(y/\mu)=1$ is observed at the $0.1$\% level for $\mu=67$ and greater.  The reader is advised to use the tables
below only for $\mu=50$ or less, where they have been verified to give correct coverage at the $1$\% level.

The reader is cautioned that these tables were generated using a unit density, uniformly distributed theoretical
function as input, and thus cannot be directly applied, as in the maximum gap case, to data to obtain an upper limit.
The data must be transformed such that it is uniformly distributed assuming a given model as its true distribution.
The necessary transformation can be found in Appendix C of~\cite{r:yellin2}.  This subtlety can be avoided by generating
a set of CDFs for every parameter change in the theoretical model considered for the data, but in most applications 
this will be prohibitively computationally intensive.

%%%%%%%%%%%%%%%%%%%%%%%%%%%%%%%%%%%%%%%%%%%%%%%%%%%%%%%%%%%%%%%%%%%%%%%%%

\begin{acknowledgments}
This work was supported by the NSF Graduate Research Fellowship program, the 
MIT Pappalardo Fellowship program, and the MIT Kavli Institute.  We wish to 
thank Steve Yellin for helpful discussions.
\end{acknowledgments}

%%%%%%%%%%%%%%%%%%%%%%%%%%%%%%%%%%%%%%%%%%%%%%%%%%%%%%%%%%%%%%%%%%%%%%%%%

\bibliography{patchPaper}% Produces the bibliography via BibTeX.

%%%%%%%%%%%%%%%%%%%%%%%%%%%%%%%%%%%%%%%%%%%%%%%%%%%%%%%%%%%%%%%%%%%%%%%%%

\begin{table*}[htbp]
\caption{Table of maximum patch CDF's (the $\hat{h}_{c,k}(x_{SI})$'s of section~\ref{sec:mc_cdf}) given an observation of $n=1,...,54$ events.  
These values of the CDFs are given as a function of the observed maximum patch divided by the total expected number of events.  Upper limits
can be set on two-dimensional data with these tabulated CDFs following the recipe laid out in section~\ref{subsec:recipe}.\label{t:tab1}}
\begin{ruledtabular}
\begin{tabular}{lccccccccccccccccccc}
(y/$\mu$) &        n=1 &        n=2 &        n=3 &        n=4 &        n=5 &        n=6 &        n=7 &        n=8 &        n=9 &        n=10 &        n=11 &        n=12 &        n=13 &        n=14 &        n=15 &        n=16 &        n=17 &        n=18 \\ \hline
    $\leq$ 0.17 &      0.000 &      0.000 &      0.000 &      0.000 &      0.000 &      0.000 &      0.000 &      0.000 &      0.000 &      0.000 &      0.000 &      0.000 &      0.000 &      0.000 &      0.000 &      0.000 &      0.000 &      0.000 \\
  0.20 &      0.000 &      0.000 &      0.000 &      0.000 &      0.000 &      0.000 &      0.000 &      0.000 &      0.000 &      0.000 &      0.000 &      0.001 &      0.002 &      0.004 &      0.006 &      0.013 &      0.022 &      0.038 \\
  0.24 &      0.000 &      0.000 &      0.000 &      0.000 &      0.000 &      0.000 &      0.000 &      0.000 &      0.001 &      0.003 &      0.008 &      0.017 &      0.030 &      0.056 &      0.093 &      0.141 &      0.192 &      0.251 \\
  0.28 &      0.000 &      0.000 &      0.000 &      0.000 &      0.000 &      0.000 &      0.003 &      0.011 &      0.024 &      0.047 &      0.081 &      0.131 &      0.188 &      0.263 &      0.339 &      0.420 &      0.483 &      0.556 \\
  0.32 &      0.000 &      0.000 &      0.000 &      0.000 &      0.002 &      0.012 &      0.034 &      0.071 &      0.119 &      0.189 &      0.271 &      0.359 &      0.447 &      0.534 &      0.606 &      0.679 &      0.735 &      0.786 \\
  0.36 &      0.000 &      0.000 &      0.000 &      0.005 &      0.024 &      0.063 &      0.121 &      0.199 &      0.292 &      0.409 &      0.509 &      0.596 &      0.676 &      0.748 &      0.796 &      0.836 &      0.878 &      0.907 \\
  0.40 &      0.000 &      0.000 &      0.007 &      0.034 &      0.091 &      0.170 &      0.283 &      0.395 &      0.505 &      0.627 &      0.706 &      0.768 &      0.828 &      0.869 &      0.902 &      0.930 &      0.956 &      0.969 \\
  0.44 &      0.000 &      0.003 &      0.030 &      0.102 &      0.204 &      0.318 &      0.464 &      0.583 &      0.679 &      0.775 &      0.842 &      0.884 &      0.920 &      0.948 &      0.965 &      0.976 &      0.986 &      0.992 \\
  0.48 &      0.000 &      0.015 &      0.083 &      0.204 &      0.345 &      0.481 &      0.619 &      0.730 &      0.814 &      0.875 &      0.924 &      0.950 &      0.969 &      0.980 &      0.987 &      0.993 &      0.997 &      0.997 \\
  0.52 &      0.003 &      0.056 &      0.171 &      0.324 &      0.491 &      0.631 &      0.749 &      0.841 &      0.895 &      0.936 &      0.965 &      0.979 &      0.986 &      0.991 &      0.996 &      0.997 &      0.998 &      0.999 \\
  0.56 &      0.017 &      0.115 &      0.276 &      0.448 &      0.628 &      0.755 &      0.843 &      0.911 &      0.943 &      0.969 &      0.983 &      0.990 &      0.993 &      0.997 &      0.998 &      0.999 &      1.000 &      1.000 \\
  0.60 &      0.042 &      0.190 &      0.400 &      0.578 &      0.732 &      0.838 &      0.911 &      0.951 &      0.973 &      0.987 &      0.994 &      0.996 &      0.997 &      0.999 &      0.999 &      1.000 &      1.000 &      1.000 \\
  0.64 &      0.080 &      0.286 &      0.518 &      0.691 &      0.817 &      0.902 &      0.954 &      0.975 &      0.987 &      0.993 &      0.997 &      0.998 &      0.999 &      1.000 &      1.000 &      1.000 &      1.000 &      1.000 \\
  0.68 &      0.134 &      0.376 &      0.625 &      0.788 &      0.891 &      0.946 &      0.975 &      0.986 &      0.993 &      0.997 &      0.998 &      0.999 &      1.000 &      1.000 &      1.000 &      1.000 &      1.000 &      1.000 \\
  0.72 &      0.194 &      0.491 &      0.726 &      0.865 &      0.944 &      0.975 &      0.990 &      0.994 &      0.998 &      0.999 &      1.000 &      1.000 &      1.000 &      1.000 &      1.000 &      1.000 &      1.000 &      1.000 \\
  0.76 &      0.276 &      0.608 &      0.820 &      0.916 &      0.975 &      0.991 &      0.997 &      0.998 &      0.999 &      1.000 &      1.000 &      1.000 &      1.000 &      1.000 &      1.000 &      1.000 &      1.000 &      1.000 \\
  0.80 &      0.368 &      0.705 &      0.877 &      0.953 &      0.989 &      0.997 &      0.999 &      1.000 &      1.000 &      1.000 &      1.000 &      1.000 &      1.000 &      1.000 &      1.000 &      1.000 &      1.000 &      1.000 \\
  0.84 &      0.471 &      0.801 &      0.935 &      0.980 &      0.996 &      0.998 &      0.999 &      1.000 &      1.000 &      1.000 &      1.000 &      1.000 &      1.000 &      1.000 &      1.000 &      1.000 &      1.000 &      1.000 \\
  0.88 &      0.581 &      0.883 &      0.969 &      0.995 &      1.000 &      1.000 &      1.000 &      1.000 &      1.000 &      1.000 &      1.000 &      1.000 &      1.000 &      1.000 &      1.000 &      1.000 &      1.000 &      1.000 \\
  0.92 &      0.717 &      0.946 &      0.991 &      0.999 &      1.000 &      1.000 &      1.000 &      1.000 &      1.000 &      1.000 &      1.000 &      1.000 &      1.000 &      1.000 &      1.000 &      1.000 &      1.000 &      1.000 \\
  0.96 &      0.859 &      0.985 &      0.999 &      1.000 &      1.000 &      1.000 &      1.000 &      1.000 &      1.000 &      1.000 &      1.000 &      1.000 &      1.000 &      1.000 &      1.000 &      1.000 &      1.000 &      1.000 \\
  1.00 &      1.000 &      1.000 &      1.000 &      1.000 &      1.000 &      1.000 &      1.000 &      1.000 &      1.000 &      1.000 &      1.000 &      1.000 &      1.000 &      1.000 &      1.000 &      1.000 &      1.000 &      1.000 \\
\hline
(y/$\mu$) &        n=19 &        n=20 &        n=21 &        n=22 &        n=23 &        n=24 &        n=25 &        n=26 &        n=27 &        n=28 &        n=29 &        n=30 &        n=31 &        n=32 &        n=33 &        n=34 &        n=35 &        n=36 \\ \hline
  $\leq$0.10 &      0.000 &      0.000 &      0.000 &      0.000 &      0.000 &      0.000 &      0.000 &      0.000 &      0.000 &      0.000 &      0.000 &      0.000 &      0.000 &      0.000 &      0.000 &      0.000 &      0.000 &      0.000 \\
  0.12 &      0.000 &      0.000 &      0.000 &      0.000 &      0.000 &      0.000 &      0.000 &      0.000 &      0.000 &      0.000 &      0.000 &      0.000 &      0.000 &      0.001 &      0.002 &      0.003 &      0.003 &      0.007 \\
  0.14 &      0.000 &      0.000 &      0.000 &      0.000 &      0.000 &      0.001 &      0.002 &      0.002 &      0.003 &      0.006 &      0.009 &      0.014 &      0.021 &      0.033 &      0.046 &      0.059 &      0.074 &      0.093 \\
  0.16 &      0.001 &      0.002 &      0.003 &      0.005 &      0.009 &      0.015 &      0.025 &      0.036 &      0.052 &      0.069 &      0.093 &      0.117 &      0.143 &      0.168 &      0.199 &      0.232 &      0.273 &      0.315 \\
  0.18 &      0.010 &      0.014 &      0.025 &      0.036 &      0.055 &      0.084 &      0.119 &      0.155 &      0.193 &      0.227 &      0.271 &      0.313 &      0.352 &      0.396 &      0.437 &      0.482 &      0.521 &      0.560 \\
  0.20 &      0.057 &      0.081 &      0.117 &      0.154 &      0.200 &      0.249 &      0.293 &      0.344 &      0.385 &      0.433 &      0.492 &      0.533 &      0.571 &      0.611 &      0.645 &      0.682 &      0.717 &      0.751 \\
  0.22 &      0.164 &      0.209 &      0.269 &      0.321 &      0.370 &      0.423 &      0.476 &      0.528 &      0.576 &      0.625 &      0.676 &      0.715 &      0.744 &      0.777 &      0.800 &      0.824 &      0.854 &      0.874 \\
  0.24 &      0.313 &      0.368 &      0.432 &      0.493 &      0.547 &      0.596 &      0.651 &      0.696 &      0.731 &      0.767 &      0.804 &      0.835 &      0.855 &      0.879 &      0.893 &      0.908 &      0.928 &      0.939 \\
  0.26 &      0.486 &      0.548 &      0.609 &      0.655 &      0.703 &      0.748 &      0.786 &      0.820 &      0.841 &      0.870 &      0.895 &      0.913 &      0.929 &      0.940 &      0.950 &      0.958 &      0.966 &      0.972 \\
  0.28 &      0.627 &      0.686 &      0.736 &      0.781 &      0.814 &      0.847 &      0.874 &      0.893 &      0.912 &      0.930 &      0.946 &      0.954 &      0.962 &      0.968 &      0.975 &      0.980 &      0.985 &      0.987 \\
  0.30 &      0.745 &      0.789 &      0.826 &      0.858 &      0.881 &      0.904 &      0.926 &      0.941 &      0.956 &      0.966 &      0.975 &      0.980 &      0.985 &      0.987 &      0.991 &      0.993 &      0.995 &      0.996 \\
  0.32 &      0.826 &      0.866 &      0.895 &      0.917 &      0.932 &      0.946 &      0.958 &      0.966 &      0.975 &      0.982 &      0.989 &      0.992 &      0.994 &      0.995 &      0.996 &      0.997 &      0.998 &      0.999 \\
  0.34 &      0.887 &      0.915 &      0.931 &      0.947 &      0.956 &      0.965 &      0.975 &      0.980 &      0.987 &      0.990 &      0.994 &      0.996 &      0.997 &      0.998 &      0.999 &      0.999 &      1.000 &      1.000 \\
  0.36 &      0.931 &      0.950 &      0.963 &      0.972 &      0.977 &      0.981 &      0.988 &      0.990 &      0.993 &      0.995 &      0.996 &      0.998 &      0.999 &      1.000 &      1.000 &      1.000 &      1.000 &      1.000 \\
  0.38 &      0.960 &      0.975 &      0.982 &      0.987 &      0.990 &      0.991 &      0.995 &      0.996 &      0.998 &      0.999 &      1.000 &      1.000 &      1.000 &      1.000 &      1.000 &      1.000 &      1.000 &      1.000 \\
  0.40 &      0.979 &      0.986 &      0.991 &      0.994 &      0.995 &      0.997 &      0.999 &      0.999 &      1.000 &      1.000 &      1.000 &      1.000 &      1.000 &      1.000 &      1.000 &      1.000 &      1.000 &      1.000 \\
  $\geq$0.42 &      1.000 &      1.000 &      1.000 &      1.000 &      1.000 &      1.000 &      1.000 &      1.000 &      1.000 &      1.000 &      1.000 &      1.000 &      1.000 &      1.000 &      1.000 &      1.000 &      1.000 &      1.000 \\
\hline
(y/$\mu$) &        n=37 &        n=38 &        n=39 &        n=40 &        n=41 &        n=42 &        n=43 &        n=44 &        n=45 &        n=46 &        n=47 &        n=48 &        n=49 &        n=50 &        n=51 &        n=52 &        n=53 &        n=54 \\ \hline
$\leq$0.08 &      0.000 &      0.000 &      0.000 &      0.000 &      0.000 &      0.000 &      0.000 &      0.000 &      0.000 &      0.000 &      0.000 &      0.000 &      0.000 &      0.000 &      0.000 &      0.000 &      0.000 &      0.000 \\
0.10 &      0.000 &      0.000 &      0.000 &      0.000 &      0.001 &      0.001 &      0.001 &      0.001 &      0.002 &      0.002 &      0.004 &      0.005 &      0.008 &      0.011 &      0.013 &      0.017 &      0.022 &      0.034 \\
0.12 &      0.010 &      0.013 &      0.021 &      0.025 &      0.034 &      0.045 &      0.057 &      0.067 &      0.084 &      0.098 &      0.118 &      0.142 &      0.167 &      0.190 &      0.214 &      0.240 &      0.267 &      0.301 \\
0.14 &      0.114 &      0.144 &      0.172 &      0.198 &      0.228 &      0.261 &      0.297 &      0.329 &      0.358 &      0.390 &      0.422 &      0.458 &      0.497 &      0.527 &      0.557 &      0.586 &      0.614 &      0.640 \\
0.16 &      0.353 &      0.396 &      0.434 &      0.469 &      0.504 &      0.540 &      0.582 &      0.617 &      0.648 &      0.674 &      0.700 &      0.724 &      0.751 &      0.775 &      0.795 &      0.813 &      0.831 &      0.847 \\
0.18 &      0.603 &      0.641 &      0.673 &      0.706 &      0.731 &      0.763 &      0.786 &      0.806 &      0.828 &      0.847 &      0.864 &      0.876 &      0.893 &      0.901 &      0.914 &      0.925 &      0.934 &      0.943 \\
0.20 &      0.783 &      0.812 &      0.834 &      0.857 &      0.876 &      0.893 &      0.909 &      0.920 &      0.932 &      0.940 &      0.948 &      0.955 &      0.960 &      0.965 &      0.970 &      0.973 &      0.977 &      0.982 \\
0.22 &      0.897 &      0.913 &      0.924 &      0.937 &      0.947 &      0.956 &      0.964 &      0.969 &      0.975 &      0.979 &      0.982 &      0.984 &      0.986 &      0.988 &      0.990 &      0.991 &      0.993 &      0.993 \\
0.24 &      0.951 &      0.959 &      0.965 &      0.973 &      0.979 &      0.982 &      0.984 &      0.986 &      0.989 &      0.991 &      0.993 &      0.994 &      0.996 &      0.997 &      0.997 &      0.999 &      0.999 &      0.999 \\
0.26 &      0.978 &      0.982 &      0.987 &      0.990 &      0.991 &      0.993 &      0.994 &      0.994 &      0.995 &      0.997 &      0.998 &      0.998 &      0.998 &      0.999 &      0.999 &      0.999 &      1.000 &      1.000 \\
0.28 &      0.990 &      0.992 &      0.995 &      0.995 &      0.996 &      0.996 &      0.997 &      0.997 &      0.998 &      0.999 &      0.999 &      0.999 &      0.999 &      0.999 &      0.999 &      0.999 &      1.000 &      1.000 \\
$\geq$0.30 &      1.000 &      1.000 &      1.000 &      1.000 &      1.000 &      1.000 &      1.000 &      1.000 &      1.000 &      1.000 &      1.000 &      1.000 &      1.000 &      1.000 &      1.000 &      1.000 &      1.000 &      1.000 \\
\end{tabular}
\end{ruledtabular}
\end{table*}

\begin{table*}[htbp]
\caption{Table of maximum patch CDF's (the $\hat{h}_{c,k}(x_{SI})$'s of section~\ref{sec:mc_cdf}) given an observation of $n=55,...,100$ events.  
These values of the CDFs are given as a function of the observed maximum patch divided by the total expected number of events.  Upper limits
can be set on two-dimensional data with these tabulated CDFs following the recipe laid out in section~\ref{subsec:recipe}.\label{t:tab2}}
\begin{ruledtabular}
\begin{tabular}{lccccccccccccccccccc}
(y/$\mu$) &        n=55 &        n=56 &        n=57 &        n=58 &        n=59 &        n=60 &        n=61 &        n=62 &        n=63 &        n=64 &        n=65 &        n=66 &        n=67 &        n=68 &        n=69 &        n=70 &        n=71 &        n=72 \\ \hline
  $\leq$0.07 &      0.000 &      0.000 &      0.000 &      0.000 &      0.000 &      0.000 &      0.000 &      0.000 &      0.000 &      0.000 &      0.000 &      0.000 &      0.000 &      0.000 &      0.000 &      0.000 &      0.000 &      0.000 \\
  0.08 &      0.001 &      0.001 &      0.001 &      0.001 &      0.001 &      0.001 &      0.002 &      0.003 &      0.003 &      0.004 &      0.005 &      0.006 &      0.008 &      0.010 &      0.012 &      0.014 &      0.017 &      0.020 \\
  0.09 &      0.006 &      0.008 &      0.011 &      0.015 &      0.018 &      0.023 &      0.029 &      0.033 &      0.041 &      0.045 &      0.053 &      0.061 &      0.073 &      0.084 &      0.097 &      0.109 &      0.125 &      0.137 \\
  0.10 &      0.041 &      0.050 &      0.067 &      0.079 &      0.092 &      0.110 &      0.127 &      0.145 &      0.162 &      0.184 &      0.205 &      0.221 &      0.244 &      0.266 &      0.290 &      0.316 &      0.348 &      0.367 \\
  0.11 &      0.156 &      0.178 &      0.205 &      0.228 &      0.249 &      0.276 &      0.299 &      0.324 &      0.354 &      0.381 &      0.409 &      0.436 &      0.460 &      0.484 &      0.512 &      0.541 &      0.567 &      0.592 \\
  0.12 &      0.327 &      0.351 &      0.383 &      0.412 &      0.436 &      0.470 &      0.500 &      0.526 &      0.552 &      0.577 &      0.600 &      0.626 &      0.655 &      0.674 &      0.699 &      0.722 &      0.738 &      0.758 \\
  0.13 &      0.504 &      0.534 &      0.570 &      0.594 &      0.616 &      0.640 &      0.665 &      0.690 &      0.713 &      0.733 &      0.754 &      0.770 &      0.791 &      0.810 &      0.828 &      0.841 &      0.853 &      0.865 \\
  0.14 &      0.669 &      0.696 &      0.722 &      0.746 &      0.764 &      0.780 &      0.800 &      0.816 &      0.836 &      0.852 &      0.864 &      0.873 &      0.884 &      0.895 &      0.911 &      0.917 &      0.924 &      0.932 \\
  0.15 &      0.781 &      0.800 &      0.822 &      0.837 &      0.854 &      0.864 &      0.879 &      0.890 &      0.904 &      0.914 &      0.923 &      0.929 &      0.936 &      0.943 &      0.953 &      0.955 &      0.959 &      0.962 \\
  0.16 &      0.859 &      0.874 &      0.890 &      0.899 &      0.911 &      0.917 &      0.926 &      0.934 &      0.943 &      0.947 &      0.956 &      0.959 &      0.963 &      0.968 &      0.975 &      0.976 &      0.979 &      0.982 \\
  0.17 &      0.915 &      0.927 &      0.936 &      0.943 &      0.950 &      0.952 &      0.958 &      0.963 &      0.968 &      0.971 &      0.976 &      0.978 &      0.980 &      0.983 &      0.987 &      0.988 &      0.989 &      0.991 \\
  0.18 &      0.950 &      0.957 &      0.963 &      0.968 &      0.971 &      0.973 &      0.978 &      0.981 &      0.986 &      0.987 &      0.988 &      0.990 &      0.991 &      0.993 &      0.995 &      0.995 &      0.995 &      0.996 \\
  0.19 &      0.974 &      0.978 &      0.981 &      0.984 &      0.986 &      0.987 &      0.990 &      0.991 &      0.994 &      0.995 &      0.995 &      0.996 &      0.996 &      0.997 &      0.998 &      0.998 &      0.998 &      0.999 \\
  0.20 &      0.985 &      0.987 &      0.989 &      0.991 &      0.993 &      0.994 &      0.995 &      0.996 &      0.998 &      0.998 &      0.998 &      0.998 &      0.999 &      0.999 &      0.999 &      0.999 &      1.000 &      1.000 \\
  0.21 &      0.991 &      0.993 &      0.993 &      0.995 &      0.997 &      0.997 &      0.998 &      0.998 &      0.999 &      0.999 &      1.000 &      1.000 &      1.000 &      1.000 &      1.000 &      1.000 &      1.000 &      1.000 \\
  0.22 &      0.995 &      0.996 &      0.996 &      0.997 &      0.998 &      0.999 &      0.999 &      0.999 &      0.999 &      1.000 &      1.000 &      1.000 &      1.000 &      1.000 &      1.000 &      1.000 &      1.000 &      1.000 \\
  0.23 &      0.998 &      0.999 &      0.999 &      0.999 &      1.000 &      1.000 &      1.000 &      1.000 &      1.000 &      1.000 &      1.000 &      1.000 &      1.000 &      1.000 &      1.000 &      1.000 &      1.000 &      1.000 \\
  $\geq$0.24 &      1.000 &      1.000 &      1.000 &      1.000 &      1.000 &      1.000 &      1.000 &      1.000 &      1.000 &      1.000 &      1.000 &      1.000 &      1.000 &      1.000 &      1.000 &      1.000 &      1.000 &      1.000 \\
\hline
(y/$\mu$) &        n=73 &        n=74 &        n=75 &        n=76 &        n=77 &        n=78 &        n=79 &        n=80 &        n=81 &        n=82 &        n=83 &        n=84 &        n=85 &        n=86 &        n=87 &        n=88 &        n=89 &        n=90 \\ \hline
  $\leq$0.05 &      0.000 &      0.000 &      0.000 &      0.000 &      0.000 &      0.000 &      0.000 &      0.000 &      0.000 &      0.000 &      0.000 &      0.000 &      0.000 &      0.000 &      0.000 &      0.000 &      0.000 &      0.000 \\
  0.06 &      0.000 &      0.000 &      0.000 &      0.000 &      0.000 &      0.000 &      0.000 &      0.000 &      0.000 &      0.000 &      0.000 &      0.000 &      0.000 &      0.000 &      0.000 &      0.001 &      0.001 &      0.001 \\
  0.07 &      0.000 &      0.001 &      0.001 &      0.003 &      0.004 &      0.005 &      0.005 &      0.007 &      0.007 &      0.010 &      0.012 &      0.015 &      0.018 &      0.021 &      0.023 &      0.027 &      0.031 &      0.039 \\
  0.08 &      0.024 &      0.029 &      0.035 &      0.042 &      0.050 &      0.058 &      0.068 &      0.078 &      0.087 &      0.101 &      0.114 &      0.128 &      0.140 &      0.156 &      0.172 &      0.186 &      0.205 &      0.222 \\
  0.09 &      0.154 &      0.173 &      0.190 &      0.206 &      0.226 &      0.246 &      0.271 &      0.295 &      0.314 &      0.342 &      0.363 &      0.386 &      0.409 &      0.429 &      0.449 &      0.472 &      0.498 &      0.518 \\
  0.10 &      0.389 &      0.414 &      0.437 &      0.458 &      0.478 &      0.502 &      0.528 &      0.552 &      0.570 &      0.593 &      0.613 &      0.632 &      0.653 &      0.668 &      0.686 &      0.703 &      0.721 &      0.737 \\
  0.11 &      0.613 &      0.634 &      0.653 &      0.672 &      0.690 &      0.706 &      0.728 &      0.743 &      0.760 &      0.775 &      0.789 &      0.804 &      0.814 &      0.824 &      0.836 &      0.847 &      0.860 &      0.869 \\
  0.12 &      0.774 &      0.790 &      0.803 &      0.818 &      0.831 &      0.843 &      0.852 &      0.861 &      0.873 &      0.882 &      0.893 &      0.902 &      0.909 &      0.916 &      0.922 &      0.927 &      0.933 &      0.940 \\
  0.13 &      0.876 &      0.890 &      0.896 &      0.904 &      0.912 &      0.920 &      0.925 &      0.930 &      0.938 &      0.944 &      0.950 &      0.956 &      0.960 &      0.963 &      0.966 &      0.969 &      0.971 &      0.975 \\
  0.14 &      0.938 &      0.946 &      0.950 &      0.956 &      0.960 &      0.964 &      0.967 &      0.971 &      0.975 &      0.977 &      0.979 &      0.981 &      0.984 &      0.986 &      0.987 &      0.988 &      0.989 &      0.989 \\
  0.15 &      0.965 &      0.971 &      0.974 &      0.977 &      0.979 &      0.981 &      0.982 &      0.985 &      0.989 &      0.991 &      0.992 &      0.993 &      0.994 &      0.995 &      0.996 &      0.996 &      0.997 &      0.997 \\
  0.16 &      0.984 &      0.986 &      0.987 &      0.989 &      0.991 &      0.992 &      0.992 &      0.993 &      0.994 &      0.996 &      0.996 &      0.996 &      0.996 &      0.997 &      0.998 &      0.998 &      0.998 &      0.999 \\
  0.17 &      0.994 &      0.995 &      0.995 &      0.996 &      0.997 &      0.997 &      0.997 &      0.997 &      0.998 &      0.999 &      0.999 &      0.999 &      0.999 &      0.999 &      0.999 &      0.999 &      0.999 &      0.999 \\
  0.18 &      0.998 &      0.998 &      0.998 &      0.998 &      0.998 &      0.999 &      0.999 &      0.999 &      0.999 &      0.999 &      0.999 &      0.999 &      0.999 &      0.999 &      0.999 &      1.000 &      1.000 &      1.000 \\
  0.19 &      0.999 &      0.999 &      0.999 &      0.999 &      0.999 &      0.999 &      0.999 &      0.999 &      1.000 &      1.000 &      1.000 &      1.000 &      1.000 &      1.000 &      1.000 &      1.000 &      1.000 &      1.000 \\
  $\geq$0.20 &      1.000 &      1.000 &      1.000 &      1.000 &      1.000 &      1.000 &      1.000 &      1.000 &      1.000 &      1.000 &      1.000 &      1.000 &      1.000 &      1.000 &      1.000 &      1.000 &      1.000 &      1.000 \\
\hline
(y/$\mu$) &        n=91 &        n=92 &        n=93 &        n=94 &        n=95 &        n=96 &        n=97 &        n=98 &        n=99 &        n=100 \\ \hline
  $\leq$0.05 &      0.000 &      0.000 &      0.000 &      0.000 &      0.000 &      0.000 &      0.000 &      0.000 &      0.000 &      0.000 \\
     0.06 &      0.001 &      0.002 &      0.002 &      0.003 &      0.003 &      0.004 &      0.006 &      0.007 &      0.010 &      0.011 \\
     0.07 &      0.045 &      0.054 &      0.060 &      0.068 &      0.075 &      0.083 &      0.095 &      0.105 &      0.116 &      0.125 \\
     0.08 &      0.238 &      0.258 &      0.275 &      0.298 &      0.317 &      0.338 &      0.358 &      0.379 &      0.399 &      0.421 \\
     0.09 &      0.536 &      0.556 &      0.574 &      0.596 &      0.611 &      0.630 &      0.647 &      0.666 &      0.680 &      0.691 \\
     0.10 &      0.751 &      0.765 &      0.778 &      0.793 &      0.801 &      0.811 &      0.824 &      0.837 &      0.848 &      0.858 \\
     0.11 &      0.877 &      0.887 &      0.893 &      0.903 &      0.908 &      0.913 &      0.919 &      0.926 &      0.929 &      0.936 \\
     0.12 &      0.944 &      0.949 &      0.953 &      0.958 &      0.959 &      0.963 &      0.966 &      0.968 &      0.970 &      0.974 \\
     0.13 &      0.977 &      0.979 &      0.981 &      0.983 &      0.985 &      0.986 &      0.988 &      0.988 &      0.989 &      0.991 \\
     0.14 &      0.990 &      0.990 &      0.991 &      0.993 &      0.993 &      0.994 &      0.995 &      0.995 &      0.995 &      0.996 \\
     0.15 &      0.998 &      0.998 &      0.998 &      0.998 &      0.999 &      0.999 &      0.999 &      0.999 &      0.999 &      1.000 \\
     0.16 &      0.999 &      0.999 &      1.000 &      1.000 &      1.000 &      1.000 &      1.000 &      1.000 &      1.000 &      1.000 \\
     0.17 &      0.999 &      0.999 &      1.000 &      1.000 &      1.000 &      1.000 &      1.000 &      1.000 &      1.000 &      1.000 \\
     $\geq$0.18 &      1.000 &      1.000 &      1.000 &      1.000 &      1.000 &      1.000 &      1.000 &      1.000 &      1.000 &      1.000 \\
\end{tabular}
\end{ruledtabular}
\end{table*}
\end{document}